\documentclass[a4paper,11pt]{article}
\pdfoutput=1 % if your are submitting a pdflatex (i.e. if you have
\usepackage{jinstpub} % for details on the use of the package, please
\usepackage{lineno}
\usepackage[misc]{ifsym} 
\usepackage{bbding}
\title{\boldmath Analysis methods for in-beam PET images in proton therapy treatment verification:  a comparison based on Monte Carlo simulations}

\author[a,b]{M. Moglioni}
\author[ a]{A.C.Kraan \Envelope\hspace{0.1mm} }\note{Corresponding author.}
\author[a,c]{, A. Berti} 
\author[a,b]{, P. Carra}
\author[d]{, P. Cerello}
\author[e]{, M. Ciocca} 
\author[d]{, V. Ferrero}
\author[d]{, E. Fiorina}
\author[a]{, E. Mazzoni} 
\author[a,b]{, M. Morrocchi}
\author[d]{, F. Pennazio}
\author[a]{, A. Retico}
\author[a,b]{, V. Rosso}
\author[a,b]{, G. Sportelli}
\author[e]{, V. Vitolo}
\author[a,b]{\textit{and} G. Bisogni}
\affiliation[a]{Istituto Nazionale di Fisica Nucleare, Sezione di Pisa, Pisa, Italy}
\affiliation[b]{University of Pisa, Pisa, Italy}
\affiliation[c]{Institute of Information Science and Technologies, National Research Council of Italy, Pisa, Italy}
\affiliation[d]{Istituto Nazionale di Fisica Nucleare, Sezione di Torino, Torino, Italy}
 \affiliation[e]{Centro Nazionale di Adroterapia Oncologica, Pavia, Italy}

\emailAdd{aafke@pi.infn.it}

\abstract{ 

\textbf{Background and purpose:} In-beam Positron Emission Tomography (PET) is one of the modalities that can be used for in-vivo non-invasive treatment monitoring in proton therapy. PET distributions obtained during various treatment sessions can be compared in order to identify regions that have anatomical changes. The purpose of this work is to test and compare different analysis methods in the context of inter-fractional PET image comparison for proton treatment verification.

\textbf{Methods:} For our study we  used the FLUKA Monte Carlo code and artificially generated CT scans to simulate in-beam PET distributions at different stages during proton therapy treatment. We compared the Beam-Eye-View method, the Most-Likely-Shift method, the Voxel-Based-Morphology method and the gamma evaluation method to compare PET images at the start of treatment, and after a few weeks of treatment. The results were compared to the CT scan.

\textbf{Results and conclusions:} Three-dimensional methods like VBM and gamma are preferred above two-dimensional methods like MLS and BEV if much statistics is available, since the these methods allow to identify the regions with anomalous activity. The VBM approach has as disadvantage that a larger number of MC simulations is needed. The gamma analysis has the disadvantage that no clinical indication exist on tolerance criteria. In terms of calculation time, the BEV and MLS method are preferred. We recommend to use the four methods together, in order to best identify the location and cause of the activity changes. 

}


\proceeding{23rd International Workshop on Radiation Imaging Detectors\\
  Riva del Garda, Italy \\
  26 – 30 June 2022}

\begin{document}
\maketitle
\flushbottom

\section{Introduction}
\label{sec:intro}
Proton therapy is a radiation therapy in which proton beams are used to destroy tumor cells. With respect to
conventional radiotherapy with X-rays and electrons, proton therapy allows for more conformal dose distributions, thanks to the characteristic depth dose profile of charged particles. %(Bragg peak)
%~\cite{loeffler}. 
However, proton therapy is more sensitive to uncertainties, that can possibly lead to dose distortions. %, if not properly taken into account. 
 Among the sources of uncertainties are anatomical changes
for instance from organ motion, tumour regression, or weight loss/gain, which can occur
during the course of treatment~\cite{paganetti, albertini}. 

Patients treated with radiotherapy for head and neck cancer frequently experience anatomical changes. For these patients a control CT is generally acquired after a few weeks of treatment (see for instance \cite{morgan, kraan}). %The scheduling of the CT scan is decided by the radiation oncologist, mostly based on clinical experience and group-based protocols. Depending on the
%outcome, the radiation oncologist may decide for treatment replanning. %Daily dose verification would
%Regarding daily 
%Cone Beam Computed Tomography (CBCT) can provide valuable information about the anatomy and setup of the patient before treatment, however it does not give indications about possible dose modifications during treatment.
In-vivo range monitoring can be a useful tool to support clinical personnel in the decision on when to schedule this control CT, which is normally scheduled according to clinical experience of the radiation oncologist. %If problems are revealed before the scheduling of the CT, it can possibly be anticipated. And vice-versa, if no issues are found, the CT scan could be postponed. 
Positron Emission Tomography (PET) is an in-vivo monitoring techniques that can be used for this scope~\cite{enghardt,parodi,nishio,knopf,kraanrev}.
Nuclear interactions of the particle beams with the patient tissue can result in the production of $\beta^+$-isotopes, like $^{15}$O, $^{11}$C, $^{10}$C. These isotopes decay emitting a positron, that annihilates into two 511 keV photons, that can be detected with a PET system. The PET activity is indirectly correlated with the delivered dose. By comparing the acquired PET images along the treatment course, it is possible to estimate whether modifications in delivered dose occur.

There are different PET data acquisition modalities, differing in acquisition time. In-beam PET are acquired inside the treatment room, allowing to obtain real-time information and to exploit short-lived isotopes~\cite{parodi, dendooven, kraanrev}. %Alternative methods to PET imaging include the detection of prompt gammas, charged energetic particles, or acoustic imaging~\cite{parodirev}.
At the National Center of Oncological Hadrontherapy (CNAO), in
Pavia, Italy, a bimodal imaging system, featuring in-beam PET and charged particle detection, is installed, called INSIDE (INnovative Solution for In-beam Dosimetry in hadronthErapy)~\cite{bisogni, doseprofiler}. %The various PET images acquired during different treatment sessions can be compared one to another in order to detect important dose distortions. 

The goal of the present study is to assess the effectiveness and compare various methods for the analysis of in-beam PET images. This will be done with the help of simulated PET images, that are based on the INSIDE in-beam PET geometry and artificially modified CT scans. Specifically, four different methods will be compared, including the Beam-Eye-View (BEV) method~\cite{fiorina2020}, the Most-Likely-Shift method~\cite{frey}%both widely used and focusing on detecting differences at the end of range of the activity distributions. Moreover, we compare  less frequently applied methods like
, the Voxel-Based-Morphometry approach~\cite{kraanvbm} and the gamma-index analysis~\cite{low}. %The comparison will be based on Monte Carlo simulations of the PET activity generated in real and synthetically generated CT scans.  %assess whether it the $\gamma$-index can be useful for intra-fractional comparisons of in-beam PET images. The gamma metric was originally developed many years ago to evaluate dose calculation algorithms and to compare dose distributions, either measured or calculated~\cite{low}. Regarding comparisons of PET activity distributions, only a few works were found ~\cite{knopf2011, zhang2018}, but both regard offline-PET and both regard comparisons between predictions and real data data, rather than inter-fractional comparisons.  %In this study we try to  % The former regarded 2-dimensional data-MC comparisons in offline-PET~\cite{knopf2011} and the latter 3D PET: measurements and simulations. %both regarding offline-PET, including awthe prescribed and simulated dose
%was used previously in a comparisonoff-line hardly used for comparing PET activity distributions. 
%In the present analysis we will use the gamma analysis to compare simulated PET images, corresponding to different treatment sessions. In particular, we investigate whether anatomical changes, artificially introduced in a CT scan, can be revealed and correctly located with the gamma analysis. This feasibility study will be based on FLUKA Monte Carlo simulations.

%A possible advantage of this method with respect to the range difference methods is the entire activity distribution is considered, rather than only the end-of-range. Moreover the gamma distribution is three-dimensional, in contrast to the BEV and MLS methods, which are two-dimensional. Finally, the method is already widely used for dose comparisons, making it easier to eventually translate the results into useful information for clinical personnel. 

%This bimodal architecture allows for the detection of annihilation photons with a PET detector, as well as for the detection of charged particles, also produced as a result of nuclear interactions. Since 2019 INSIDE is under clinical trial. During this trial (ClinicalTrials.gov ID: NCT03662373), we acquired IB-PET
%The data analysis of eight patients, based on the most widely used analysis methods to detect inter-fractional range differences, was recently published.  analysis methods Beam that received proton therapy treatments, fractionated in 6 weeks (about 30 sessions) along the entire course of their treatment. The first phase of the trial was completed in March 2020.

\section{Methods and Materials}
\subsection{FLUKA MC simulations}
\begin{figure}[!ht]
    \centering
    \includegraphics[trim=0 50 0 50, clip, width=0.95\textwidth]{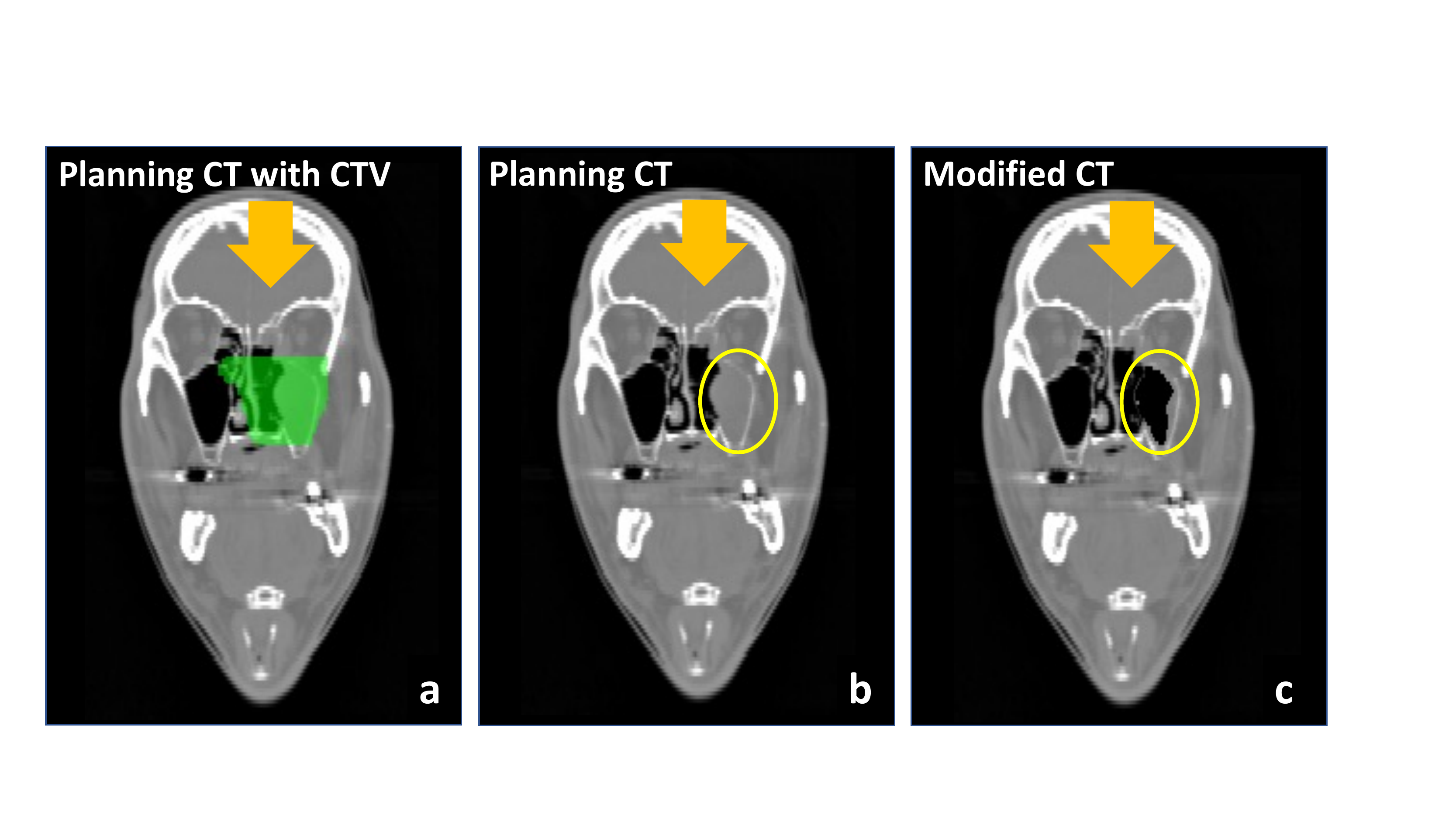}
    \caption{(a) Slice of the planning CT of the SCC patient, with the CTV indicated in green. (b) The same, but with region of interest indicated as yellow circle, containing the sinonasal cavity.
(c) The same slice, but now including an artificially introduced anatomical change. The emptying of the sinonasal cavity is clearly visible.}
    \label{fig:ct}
\end{figure}

For our study we used a CT scan of a 70-year-old patient treated in 2018 for Squamous Cell Carcinoma
(SCC) with proton therapy at CNAO. This patient was used in previous studies~\cite{kraanvbm, fiorina2020} and only the most relevant information is reported here. A planning CT scan %with resolution 0.98 x 0.98 x 2 mm$^3$ 
was available with corresponding structure set. In Figure~\ref{fig:ct}(a) a slice of the planning CT scan is given, where the Clinical Target Volume (green zone) is shown. Figure~\ref{fig:ct}(b) shows the same slice, with the sinonasal cavity region highlighted with a yellow circle. The tumor region received 2 Gy per treatment session (66 Gy in 33 fractions). An intensity modulated proton therapy
(IMPT) treatment plan was available for this CT scan. We introduced an anatomical change in the patient CT, simulating a cavity emptying~\cite{kraanvbm},   by modifying the Hounsfield units, as can be seen in Fig.~\ref{fig:ct}(c). The total volume that we changed was 13 ml. %Thus, we had two CT scans: the original CT scan, and the CT scan including the emptying of the cavity, 
Such a volume change can realistically occur over weeks of treatment.

%gamma >1 looks larger changed volume because there are more slices in which gamma is present, CTV low is presebt but not the changed volume by Andrea.  
 
\begin{figure}[!h]
    \centering
    \includegraphics[trim=0 50 0 100, clip, width=0.95\textwidth]{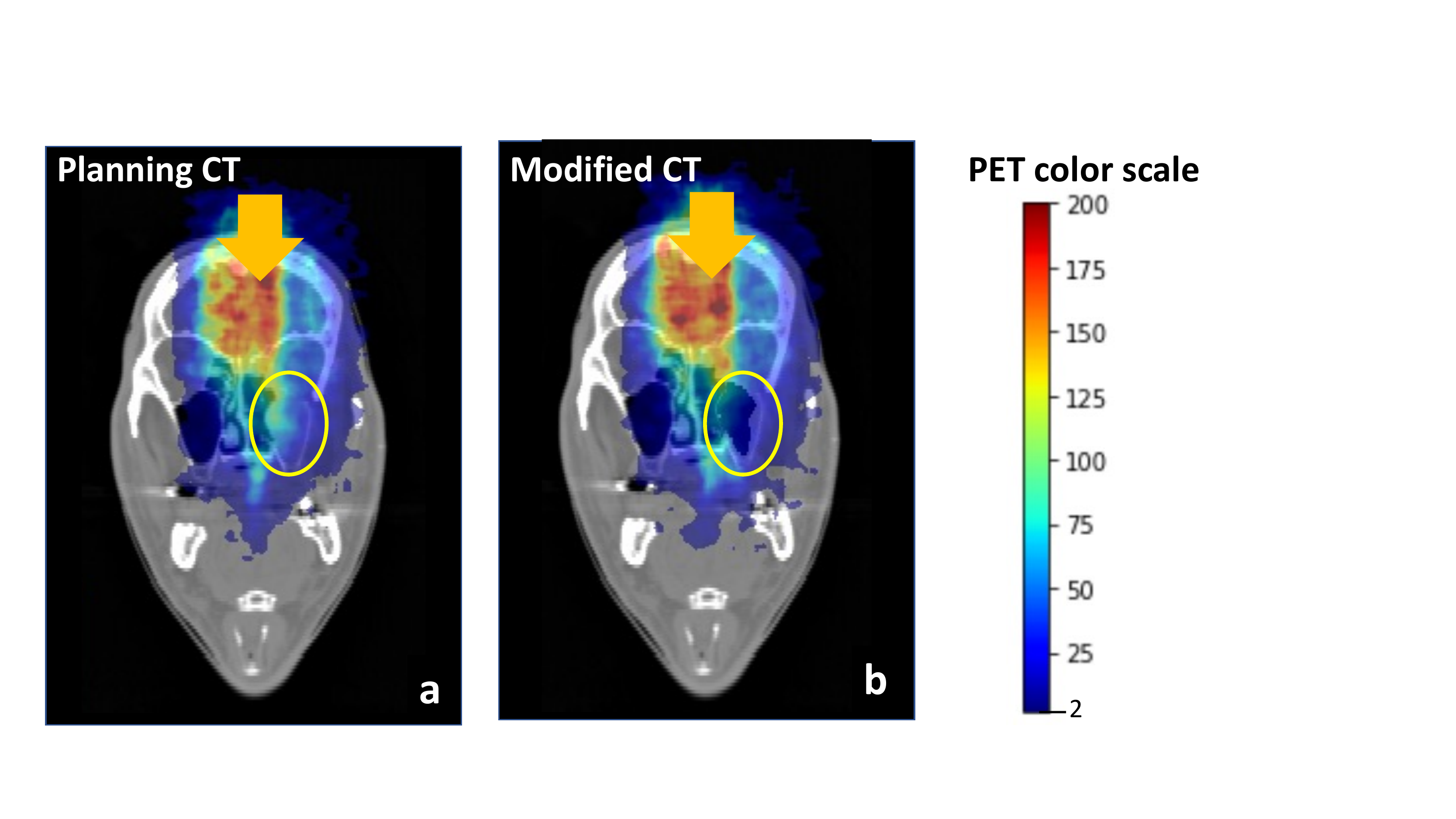}
    \caption{(a) Slice of the planning CT of the SCC patient together with the simulated PET distribution.
(b) Same but now including an artificially introduced anatomical change. }
    \label{fig:ctpet}
\end{figure}

For simulating the PET activity, a tool previously developed for INSIDE~\cite{pennazio, fiorinaMC} was used, based on FLUKA Monte Carlo simulations~\cite{fluka}. Shortly summarizing, it included the simulation of the CNAO beamline, the pencil beam transverse dimension and shape, the time structure of the delivered protons, and the patient geometry, obtained by importing the CT scan in a voxel FLUKA geometry. Radiation transport and interactions of protons in the patient were simulated, as well as the generation and decay of the $\beta^+$ emitting isotopes. %We extracted the spatial coordinates and production time of the $\beta^+$emitting isotope 
%($^{11}$C,$^{15}$O, and so on).
 %This collection of produced isotopes is referred to as the isotope production map (IPM). The annihilation decay and transport of the 511 keV photons in the patient was simulated. Also, 
 The geometry of the INSIDE PET detector was included, with the pixelated LSF crystals and the energy depositions of the photons. We selected coincidences, that occurred between the start of treatment, in between the beam spills, up to about 6 minutes. The lines-of-responses (LORs) were created and subsequent image reconstruction was done with an iterative maximum likelihood estimation maximization (MLEM) procedure. In the end, a 3-D PET image was obtained, with a field-of-view (FOV) of $22.4 \times 11.2 \times 26.4$ cm$^3$, with $1.6\times 1.6\times 1.6$ mm$^3$ voxels. The obtained images suffered from reconstruction artefacts in the direction perpendicular to the two PET detector planes, a well-known problem for planar PET scanners. We applied a median filter of 1.6 mm (1 voxel) in all directions in order to decrease statistical fluctuations. The PET activity was simulated for the original and the modified CT scan, yielding a PET activity distribution without (Figure~\ref{fig:ctpet} a) and one with (Figure~\ref{fig:ctpet} b) anatomical changes. These are hereafter referred to as $PET_{ref}$ and $PET_{mod}$, respectively. 

\subsection{Analysis methods}
To compare the two PET images of Fig.~\ref{fig:ctpet}, various analysis techniques have been developed in the past. In this work we compare four different methods:
\begin{itemize}
%    \item 1-D range analysis, as used for instance in early in-beam PET studies~\cite{rosso, kraanrange}. For a given PET image a 1-D activity distribution along the beam axis (z) is obtained by summing  all activity in the voxels with |$x^2$ + $y^2$| < r cm) and plot it $z$.  %The 50\% rise and 50\% fall-off of the 1-D profile positions can than be determined. 
%The difference between them, $\Delta w_{50\%}$ can be evaluated and is defined as the range. The difference in range b®etween the original and the modified PET image is defined as $\delta R_{1-D}$. %This method was mostly applied for irradiation of small volumes. 
    % discussion: method mostlty for single pencil beams

    \item The Beam-Eye-View method~\cite{fiorina2020}.
    This method focuses on the distal part of the activity (see Figure~\ref{eor} a). 
    \begin{figure}[!t]
    \centering
    \vspace{-0.5cm}
    \includegraphics[trim=10 46 0 230, clip, width=1.1\textwidth]{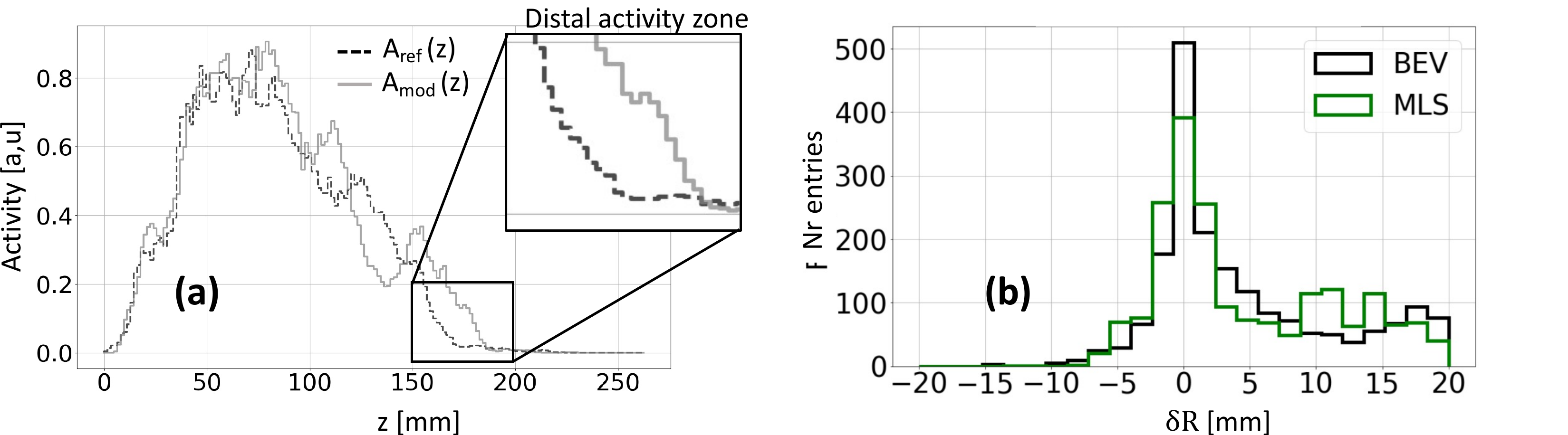}%\includegraphics[trim=-100 10 0 0, clip, width=0.5\textwidth]{deltaR.png}
    \caption{\label{eor} (a) Example of a reference PET image profile ($A_{ref}(z)$, black dashed) and the corresponding modified PET profile ($A_{mod}(z)$, grey solid) in point $(x,y)=(35, 70)$. The BEV and MLS methods focus on the very last part of an activity profile. (b) The distribution of range differences found between $PET_{ref}$ and $PET_{mod}$ for all ($x,y$) points for the BEV (black) and MLS (green) method.\label{methods} }
\end{figure}
This method is based on a multi-threshold approach to extract range differences. %Thresholds from 2\% up to 8\% on the maximum of the entire PET image at 0.5\% steps were considered. Briefly 
    For each pair of coordinates $x,y$ in the transverse plane, an activity profile along $z$ of the reference image was constructed, and its integrated activity was determined. The maximum value of all activity profiles was evaluated, and all profiles with integrated activity less than 30\% of this value were excluded. For the remaining $(x,y)$ coordinates, we determined the $z$ coordinate where the activity value was above a threshold $t$, $R_{ref}^t$. The value of $R_{ref}^t$ was determined for thresholds from 2\% up to 8\% of the maximum of the entire PET image, with 0.5\% steps. The same was done for the corresponding profiles of the modified image. For a given $x,y$ pair, the range shift between the two profiles, $\delta R{_{\text{BEV}}}(x,y)$, was defined as: 
    
\begin{equation}
        \delta R{_{\text{BEV}}} (x,y) = \frac{1}{N}\sum_{t=1}^N \delta R{_{\text{BEV}}^t}(x,y)= \frac{1}{N}\sum_{t=1}^N\left(R_{mod}^t(x,y)-R_{ref}^t(x,y)\right)
\end{equation}
where $N$ is the total number of thresholds considered, which was 13,  $R^t_{mod}$ and $R^t_{ref}$ are the range values of the modified and the reference profile, respectively, and $\delta R^t_{BEV}(x,y)$ is the difference between them. %These are the maximum $z$-value where the activity is above threshold $t$, and $R^t_{BEV}$ is the difference between them. Thresholds were included from 2\% up to 8\% of the maximum of the entire PET image, with 0.5\% steps. 

%For coordinate $(x,y)$, the value of the range difference $\delta R_{BEV}(x,y)$ was defined as the average over the thresholds:
Then a map $ O_{BEV}(x,y,z) $ was created by filling each grid point $(x,y,z)$ as: 
\begin{equation}\label{bev_eq}
    \begin{cases}
        O_{_{\text{BEV}}} (x, y, z) = \delta R_{\text{BEV}} (x, y) &    \text{ if }
        z \le R_{8\%}^{mod}(x.y) \text{ and } | \delta R_{BEV}(x.y)| \ge 5 \text{ mm }
        \\
        O_{_{\text{BEV}}} (x, y, z) = 0 \qquad & \text{ otherwise}
    \end{cases}
\end{equation}
This map was superimposed to the CT images. The highlighted zones represent the zones where activity is present (larger than 8\% of the maximum value) and where a range difference larger than 5 mm was found. %They highlight the entire region along a pencil beam path that may be affected by a range shift larger than 5 mm. 
Positive and negative values for $\delta R$ indicate an overshoot (displayed with red) and undershoot (displayed with blue), respectively, with respect to the reference situation. 

    \item  Most-Likely-Shift method. This is also a two-dimensional range analysis, evaluating range differences in the $(x,y)$ plane, originally proposed by Frey et al~\cite{frey}. Considering two PET images that must be compared, we developed an algorithm, that provided for each $x,y$ pair an optimal shift distance along $z$, called $\delta_{MLS}$, between the PET images. Exactly the same $(x,y)$ profiles were included as in the BEV analysis. The range difference $\delta R_{MLS}$ is the $\delta_{MLS}$ value that minimizes the absolute differences in the distal part %($z_{MLS}\le z\le z_{max}$) 
    of the two activity depth profiles %, shifted against each other
    ~\cite{frey}: \begin{equation}
    \delta R_{MLS}(x,y) = \arg\min_{\delta_{MLS}} \left(\sum_{z_{MLS}}^{z_{max}} |A_{mod}(x,y,z-\delta_{MLS})-A_{ref}(x,y,z) |\right)
\end{equation}
with $A_{ref}$ and $A_{mod}$ corresponding to the reference activity value and that of the modified activity value, respectively.
 Then a three-dimensional map $ O_{MLS}(x,y,z)$ was defined as:
\begin{equation}\label{mls_eq}
    \begin{cases}
        O_{_{\text{MLS}}} (x, y, z) = \delta R_{\text{MLS}} (x, y) & \text{if } z\le z_{max}\text{ and } | \delta R_{MLS}(x.y)| \ge 5 \text{ mm }
        \\
        O_{_{\text{MLS}}} (x,y,z)=0 &\text{ otherwise}
    \end{cases}
\end{equation}  
These maps were re-oriented on the patient's CT reference frame. The interpretation of the maps is just like for the BEV method.

    \item The Voxel-Based-Morphometry method~\cite{kraanvbm}. This analysis is a three-dimensional voxel-by-voxel analysis. For each voxel of the $PET_{mod}$ image, a statistical analysis is done, testing the compatibility of the voxel value with the null hypothesis (situation without morphological changes). Rejection of the null hypothesis implies that the voxel intensity has significantly changed. 
    %The method is illustrated in Fig.~\ref{fig:vbm}. 
    %\begin{figure}[!ht]
    %\centering
    %\includegraphics[trim=0 40 0 10, clip, width=0.7\textwidth]{vbm.pdf}
    %\caption{Illustration of the voxel-based morphometry method. }
    %\label{fig:vbm}
%\end{figure}
    For this purpose, not one distribution $PET_{ref}$ must be obtained, but a large number of reference distributions must be obtained. This was done by simulating 120 times the reference PET image, each time with a new random seed, as documented in an earlier work of our group~\cite{kraanvbm}. Let us define $v_{ref}$ as the intensity value in a certain voxel in the reference PET image,  $\bar{v}_{ref}$ as the average intensity value over all $N$ replicates of the reference, and $v_{mod}$ as the intensity value in the modified PET image in the same voxel.  The empirical $p$-values in that voxel is: 
\begin{equation}\label{egp}
    p = \frac{r+1}{N+1},
\end{equation}
where $N$ was the number of replicates (120) and $r$ was the number of replicates that produced a test statistic of either $ v_{ref} \geq v_{mod}$ if $ v_{mod} \geq {\bar v}_{ref} $, 
or $ v_{ref} \leq v_{mod}$ if $ v_{mod} < {\bar v}_{ref} $. The resulting $p$-values evaluated for each voxel of the modified PET image can be reported as 3-D voxelized  maps. As significance threshold we chose  $p\le 0.025$. Significantly more and less activity in a certain voxel is indicated by purple and cyan, respectively.  %(red) and $ v^i_{obs_j} \leq {\bar v}^i_{sim} $ (blu when significantly less activity than expected), respectively.  
These colored three-dimensional maps can be overlaid onto the original CT image.%, allowing directly to localize the areas where an anomalous amount of activity was found. 

    \item The $\gamma$-index analysis. The gamma test~\cite{low} is a metric that is mostly used for dose comparisons. 
    %\end{itemize} 
 %was introduced to overcome problems related to calculations of dose difference (DD) and distance-to-agreement (DTA) calculations, typically failing in high and low gradient regions, respectively. the gamma-index combines them both into an abstract metric resembling a distance.
 When an evaluated dose distribution $D_e$ is compared to a reference dose distribution $D_r$, given at grid points $\vec{r}_e$ and $\vec{r}_r$, respectively, the gamma index $\gamma(\vec{r}_e)$ is given by:

\begin{equation}\label{gamma}
        \gamma\left(\vec{r}_{e}\right) = \min\left[\Gamma\left(\vec{r}_{e},\vec{r}_{r}\right)\right]\quad \forall\ \vec{r}_{r}
    \end{equation}
    where
    \begin{equation}\label{Gamma}
        \Gamma\left(\vec{r}_{e},\vec{r}_{r}\right)=\sqrt{\frac{\Delta r^2\left(\vec{r}_{e},\vec{r}_{r}\right)}{\text{DTA}^2}+\frac{\Delta D^2\left(\vec{r}_{e},\vec{r}_{e}\right)^2}{\text{DD}^2}}
    \end{equation}
Here $\Delta r(\vec{r}_e,\vec{r}_e)=|\vec{r}_e-\vec{r}_r$| is the distance between $\vec{r}_e$ and $\vec{r}_r$, $\Delta D(\vec{r}_e,\vec{r}_r)=D(\vec{r}_e)-D(\vec{r}_r)$ is the value of the dose difference, and the parameters $DTA$ and $DD$ are the acceptance criteria for distance-to-agreement  and dose-to-agreement respectively. %In other words, the evaluated point with the smallest $\Gamma$  corresponds best to the reference point and is referred to as the $\gamma$ index in that grid point. 
If $\gamma\le 1$, the point is accepted (compatible distributions in that grid point), and if $\gamma$ > 1 the point is rejected (incompatible distributions in that grid point). 
The dose tolerance is usually a percentage of prescribed dose (global gamma index). %Regarding the dose tolerance $DD$, this can be calculated relative to a \emph{global} value like the maximum dose value or prescribed dose value, or relative to the \empg{local} value expected in $\vec{r_r}$. Most widely applied is the global gamma-index evaluation. There is third parameter which needs to be defined, which is the dose threshold $T$. Below this threshold, the gamma-index is not calculated because it is considered irrelevant. 
In dose comparisons, widely used parameters for $DTA$ and $DD$ acceptance criteria are 3 mm and 3\% of the maximum dose, respectively~\cite{low, aapmtg}. 
In this work, we applied the gamma index analysis directly on the activity distributions, rather than on the dose. This allows to compare, as in the other proposed methods, inter-fractional IB-PET images (see discussion).  %Furthermore a threshold must be chosen, which is in dose comparisons widely varying and usually around 10\%. 
%, but higher and lower thresholds are also widely reported. As dose threshold, 10\% of the maximum dose is usually applied in dose comparisons\cite{aapmtg,gammathreshold}, but again a variety of thresholds has been used. Finally, the gamma-index passing rate, $PR$, is defined as the percentage of evaluated points satisfying the condition of gamma index $<1$. 
%\end{itemize}
%In the works regarding PET image comparisonsa global gamma analysis with tolerances of 3 mm and 45\% was used for data-simulation comparisons~\cite{knopf2011}, while tolerances of 3\% and 3 mm were used in~\cite{zhang2018}. No indications or recommendations for PET image comparisons are found in literature. We therefor empirically test several tolerances and thresholds, and investigate whether the gamma index can be used to reveal differences between the reference PET distribution $PET_{ref}$ and the modified PET distribution  $PET_{mod}$. We study:
%\begin{itemize}
%    \item Threshold values in the range from 1\% to 10\%.
%    \item Activity tolerances $DD$ ranging from 3\% to 7\%. 
%    \item Distance tolerances $DTA$ ranging from  3 mm to 7 mm. 
%    \item Local versus global $\gamma$-index calculation.
%\end{itemize} 
 We made use of the three-dimensional $\gamma$-index function implemented in python3~\cite{wendling}. 
\end{itemize}

\section{Results}

In figure~\ref{methods} b we show the distribution of $\delta R$ values that was obtained when considering all $(x,y)$ points. For both methods, an asymmetric distribution is seen towards positive values, indicating that the range value of the modified PET image is larger than that in the reference situation. In other words, we expect that there is somewhere a range overshoot in the modified PET image with respect to the reference image. This was expected: the emptying of the cavity, were liquid material is replaced by air. results in a range increase (overshoot) of the pencil beams that cross the cavity. %As a result, activity is generated beyond this region, and a positive range difference is the result. 

\begin{figure}[t!]
\centering
\vspace{-0.4cm}
\includegraphics[trim=30 70 10 20, clip, width=0.9\textwidth]{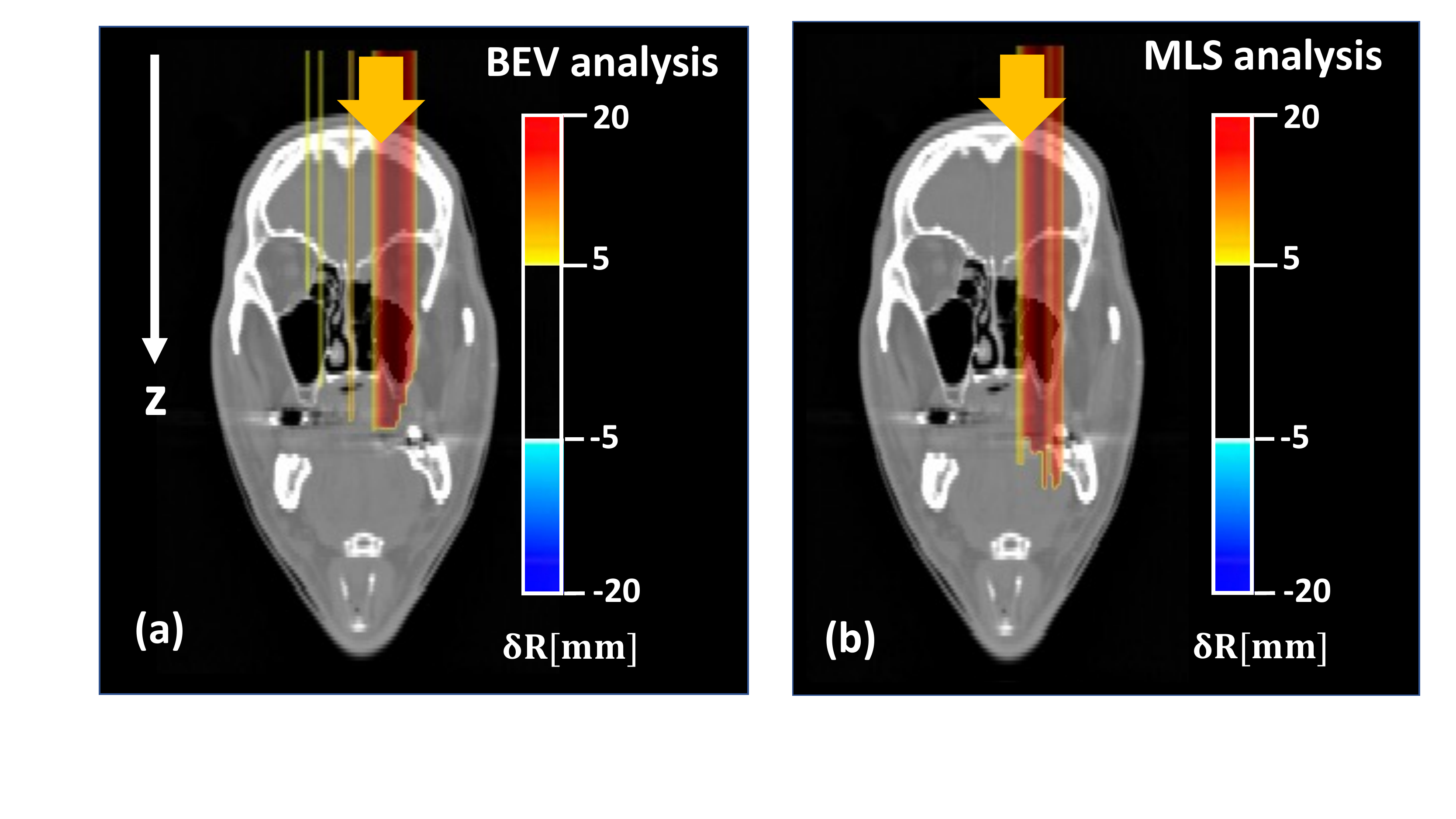}
\caption{\label{fig2d}{The maps $O_{BEV}$ (a) and $O_(MLS)$ representing the estimated zones where the pencil beams pass that are affected by a range difference between $PET_{ref}$ and $PET_{mod}$. The red color indicates that there is a beam overshoot in the modified PET image with respect to the reference. The beam direction is indicated with a yellow arrow.  \label{figmaps}}}
\end{figure}

In Figure~\ref{figmaps} we show the $O_{BEV}$ (a) and $O_{MLS}$ (b) distribution resulting from comparing $PET_{ref}$ with $PET_{mod}$, overlaid on the modified CT image. % In the green zones, no range differences are expected ($\delta R=0$). 
The red zones represent the location of the activity profiles along $z$, that have resulted in positive range differences, i.e., a beam overshoot, as a result of the emptied cavity. %This was expected in this zone, as explained above.

\begin{figure}[h!]
\begin{center}
\includegraphics[trim=30 120 0 0, clip, width=0.9\textwidth]{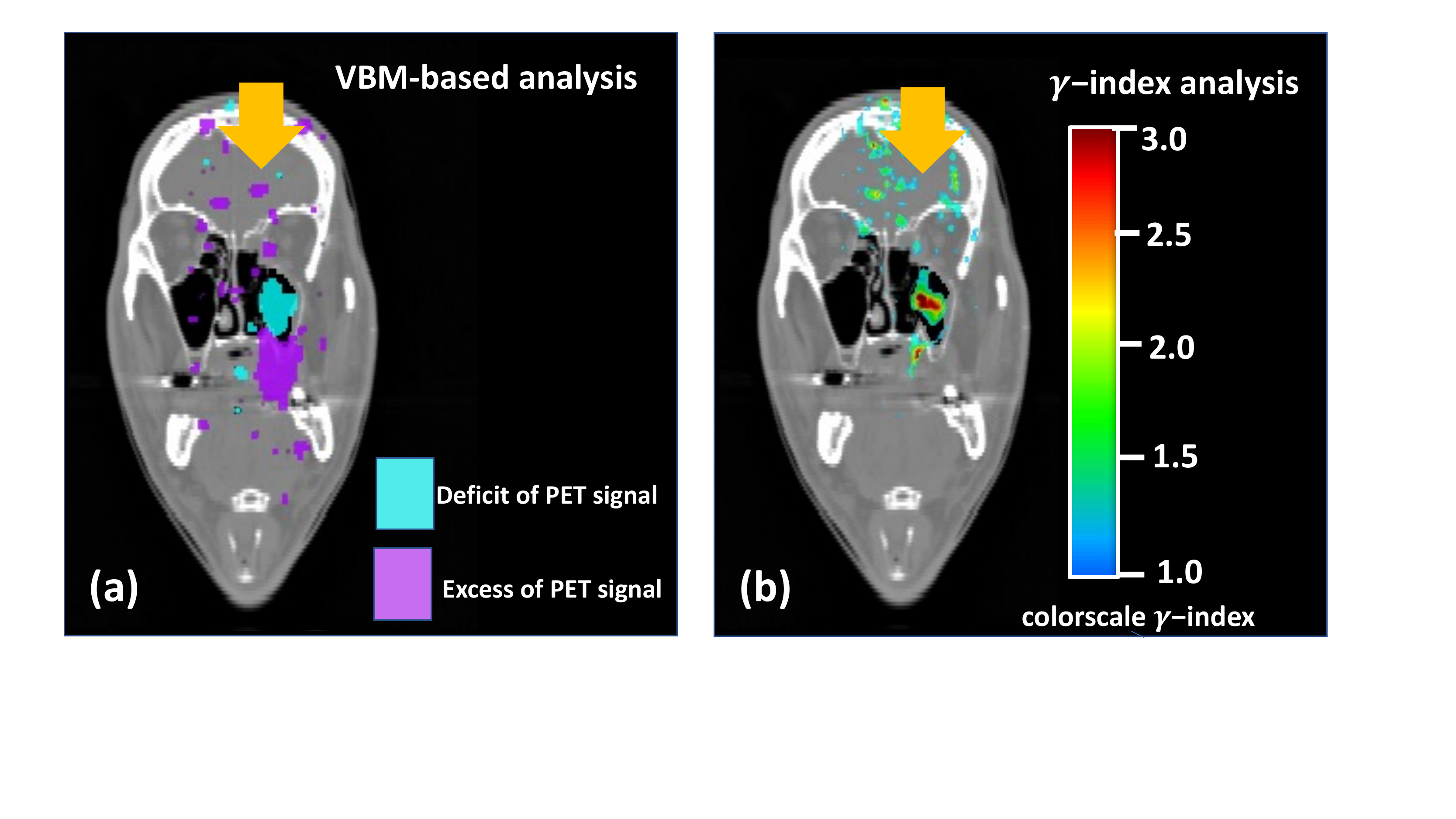}
\end{center}
\caption{(a) Example of a CT slice with superimposed the VBM maps, with the zones exhibiting a deficit (cyan) and excess (purple) or activity with respect to expectations.  (b) Example of a CT slice with superimposed the $\gamma$-index map (3 mm/3\%.) The threshold here was 5\% of the maximum activity value. In the colored regions, discrepancies were found between the reference and the modified image. The beam direction is indicated with a yellow arrow. \label{gammaimg} }
\end{figure}

In Figure~\ref{gammaimg} a we show the result of the VBM analysis. Here the purple and cyan zone represent zones, where significantly more and significantly less activity was found  in the modified PET image with respect to the reference image, respectively. We observe that the cyan region corresponds well to the air cavity. The purple region is the zone where significantly more activity was found with respect to what was expected. This a result of the beam overshoot. 

Regarding the $\gamma$-index analysis, we show in Figure~\ref{gammaimg} b the result of the analysis with tolerance criteria 3 mm/3\%. %, typical dose comparison values. 
Only gamma-index values above 1 are displayed, so that only the regions where $PET_{ref}$ and $PET_{mod}$ are incompatible are highlighted. First of all, we observe that the emptied volume is clearly highlighted with these tolerance criteria. Second, we see some very small highlighted regions in zones where no morphological changes occur. Third, the region of the beam-overshoot, is seen to be somewhat smaller than the region identified with the VBM analysis (purple zone). This has to do with the parameters in the $\gamma$-index analysis (see Section~\ref{dis}). 

%If the threshold is large (10\%), the gamma-index is not calculated in this region, since the associated activity is close to zero in $PET_{ref}$ and below threshold. When the threshold decreases, more voxels are included in the gamma-index calculation, and the overshoot can be clearly revealed. Thus, we believe that for PET distributions, a low threshold such as 1\% is more appropriate than large values like 10\% or 50\% as recommended for dose comparisons~\cite{aapmtg}. 

%\subsection{Global and local $\gamma$-index calculation}
%The local and global criteria were compared for  tolerance criteria of $DD=$3\% and a threshold of 1\% of the maximum expected activity, Figure~\ref{fig4} shows the resulting gamma-index distributions, with in (a) and (b) the global and local gamma-index calculation, respectively. We observe that the regions where incompatibilities are found are mostly similar, but the global calculation results generally in higher $\gamma$-index values for the same tolerance criteria and threshold, i.e., a lower compatibility. Also, the overshoot is much more pronounced. This is both because in regions with low activity, a local difference of 3\% is very easy to obtain due to statistical fluctuations. This is the case in the region of the overshoot, and also at the sides of the activity distributions. There is in general more noise in the local calculation. Therefore, we believe the global $\gamma$-index calculation is preferred.

%\subsection{Tolerance criteria}
%Figure~\ref{fig5} shows an example of the gamma-index distribution for different tolerance criteria, while including a threshold of 1\%. We note that with increasing tolerances, the highlighted zone decreases in size and intensity. Also, the noise dissapears when the threshold increases. This is logical, as the two PET distributions are increasingly compatible when the tolerance grows, 
%\begin{figure}[h!]
%\begin{center}
%\includegraphics[trim=0 0 0 0, clip, width=0.95\textwidth]{FigGammaCriteria.pdf}
%\end{center}
%\caption{\label{fig1}{The comparison between a few different tolerance criteria for the global $\gamma$-index calculation and $TH=1$\%: $DD=3\%$, $DTA=3$ mm (a), $DD=4\%$, $DTA=4$ mm (b), $DD=5\%$, $DTA=5$ mm (c) and $DD=7\%$, $DTA=7$ mm (d). \label{fig5} }}
%\end{figure}

\section{Discussion\label{dis}}
We investigated four different methods on the same Monte Carlo simulated patient treatment. Advantages and disadvantages of the methods are the following. 

%The 1-d activity summation method is simple, but does not give much spatial information. It is therefore more suited for single pencil beam irradiations.  

In the BEV method the range differences between two images are calculated, based on the distal part of the activity along the beam axis ($z$). No information in the rest of the profiles is used. The maps allow to reveal a beam overshoot or undershoot. The information is 2- dimensional: $\delta R$ in the ($x,y$) plane. Advantages of this method are that it is conceptually simple and the information is direct. The validity of this method was tested on data~\cite{fiorina2020}. 

The MLS method gives, just like the BEV method, 2-dimensional information. With respect to the BEV method,  the MLS method is much more complex, while giving more or less the same information. On the other hand, it is be somewhat less sensitive to fluctuations in the values in distal activity, since it does not look at a threshold value, but tries to find the best matching $\delta R$ value for the end-of-range profile. This method is being tested on data, as reported in~\cite{moglionikraan}. 

The VBM method is a 3-dimensional methods, and can possibly provide more spatial information. It displays precisely where the excess or lack of activity is located in appropriate color codes. This is interesting from a clinical point of view. The drawbacks of the method are the large number of simulations that are needed to be able to calculate voxel-by-voxel the the p-value. Also, it is a MC driven method as present, and it was only validated on MC simulations.  

The $\gamma$-index analysis is also a 3-dimensional method, indicating where the activity of a given distribution is anomalous with respect to another distribution, but does not indicate whether there is a lack or excess of activity. Moreover, the excess of activity is not clearly revealed. We believe that an optimization of the free parameters in the analysis can change this, and this is part of future research. Eventually, this method could help in transforming the activity distributions into a dose distribution~\cite{aiello,remmele}. A disadvantage of the $\gamma$-index analysis is the long calculation time, and the choice of the parameters, for which no recommendations exist for PET image comparisons.  %More work is needed to study the tolerance criteria and to determine the most appropriate threshold. %Also the method should be applied to data. 

\section{Conclusion}
The four methods presented above are all valuable methods for comparing in-beam PET images taken during subsequent fractions in proton therapy. On one hand, the VBM and $\gamma$-index analysis methods can provide 3-dimensional information about discrepancies, which cannot be done with the BEV and MLs methods. On the other hand, they suffer more from falsely identified voxels, and their calculation time is much longer. It would be beneficial, to combine the various methods, so that regions that seem to have changed in one method, are confirmed with the other methods. 

\end{document}